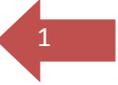

# Single-shot all-optical magnetization switching in in-plane magnetized magnetic tunnel junction

S. Geiskopf[1], J. Igarashi[1,2], G. Malinowski[1,*], J.-X. Lin[1], J. Gorchon[1], S. Mangin[1], J. Hohlfeld[1], D. Lacour[1] and M. Hehn[1,*]

[1] Université de Lorraine, CNRS, IJL, F-54000 Nancy, France
[2] National Institute of Advanced Industrial Science and Technology, Tsukuba 305-8563, Japan

[*] michel.hehn @univ-lorraine.fr, gregory.malinowski@univ-lorraine.fr

## Abstract

Single pulse All Optical Helicity-Independent Switching is demonstrated in an in-plane magnetized magnetic tunnel junction. A toggle switching of the 2nm thick $Co_{40}Fe_{40}B_{20}$ soft layer could be achieved by exchange coupling the $Co_{40}Fe_{40}B_{20}$ with a 10nm thick $Co_{85}Gd_{15}$ layer monitored by measuring the Tunnel magneto resistance of the device. The use of in plane magnetized electrodes relaxes the constrains linked to perpendicular magnetic anisotropy systems while achieving a tunneling magnetoresistance (TMR) ratio exceeding 100%. The influence of the upper electrical electrode, which is opaque to the laser beam in this study, is also discussed.

## Introduction

Magnetic Random Access Memories (MRAM) based on the use of magnetic tunnel junctions (MTJ) are at the heart of current developments in non-volatile storage media [Hos05] [Kaw12] [Bra12] [Ken15] [Apa16] [Die20]. Over time, the writing schemes for MRAM have evolved transitioning from magnetic field-based writing [Tan20] to current induced writing, either spin-transfer-torque (STT) [Slo96] [Ber96] [Ral08] [Khv13] or spin orbit torque (SOT) [Mir11] [Kri22] [Man19]. Due to its high tunnel-magnetoresistance-ratio (TMR) effect that improves readability, all the MRAM architectures are nowadays based on the use of CoFeB/MgO/CoFeB structures [Yua04] [Par04]. However, achieving TMR ratios above 100% requires annealing at high temperatures, which imposes constraints on the perpendicular magnetic anisotropy (PMA) used to reduce critical writing currents [Man09]. To overcome these challenges, complex Synthetic Antiferromagnets (SAF) for both the reference and the soft layers [Yos13] and/or associations with MgO or Pt layers [Ike10] [Got15] are often needed to get PMA while the crystallization of CoFe through the diffusion of B away from the CoFe/MgO interface requires the use of a Ta, W or Mo absorbing layers [Zhu12] [Hon17] [Liu14]. This well engineered material stack enables magnetization full reversals, and so storage of one information bit in one MRAM cell, on a time scale of the order of few ns with energy consumption of a few pJ [Yaz22].

The all-optical-switching (AOS) technology enables further acceleration of the magnetization-reversal process with sub-picosecond crossing zero magnetization times without any applied current or field. Single-pulse AOS was first discovered in GdFeCo ferrimagnetic systems [Sta07], and then extended to many Gd-based alloys or Gd/FM bilayers, where FM is a ferromagnetic layer [Rad11] [Gwe24] [Zha24] [Bee19] to MnRuGa ferrimagnetic Heusler alloys [Ban20], to Tb/Co multilayers [Sal23], and Tb/Fe and the Tb32Co68/Co/Tb32Co68 trilayer [Pen23]. In this plethora of materials, in terms of reversal speed, the fastest switching were obtained using Gd-based materials, with sub picosecond switching times with write/erase cycles of less than 10 ps [Set22]. While all those materials have PMA, single pulse AOS was also extended to in plane magnetized ferrimagnets [Ost12][Lin23] which shows single pulse AOS for a broader range of alloy concentrations [Lin23].

In view of the major impact this technology could have on MRAMs, materials hosting AOS have been incorporated into MTJs. The initial demonstration of AOS operation was presented by Chen et al. [Che17], reporting a TMR of 0.6% with CoFeGd directly interfacing the MgO tunnel barrier. Subsequently, Wang et al. [Wan22] reported AOS reversal in a junction, achieving a significantly improved TMR of 34.7%. In this case CoFeB was in contact with MgO and exchange biased to a Co/Gd bilayer. Finally, the increase in the TMR up to 74% was possible by the use of [Tb/Co]$_N$ multilayer coupled to a FeCoB which keeps PMA even at 250°C [Sal23]. However, since [Tb/Co]$_N$ multilayer shows a laser induced precessionnal switching [Sal23][Peng23], the reversal time is 100 times slower than in Gd based magnetic soft electrode.

In this work, we show that all these limitations can be relaxed in in-plane magnetized MTJs exhibiting both single pulse all-optical switching and a TMR larger than 100%. Moreover, we show that the annealing temperature can be brought to 300°C without impacting the magnetic properties.

## Experimental details and results

Conventional magnetic tunnel junctions were prepared using ultra-high vacuum sputtering with a base pressure below $5 \times 10^{-9}$ mbar. All layers were grown at $5 \times 10^{-3}$ mbar with deposition rate of the order of 0.5 Å/s (except for MgO which deposition rate is around 0.015 Å/s). The multilayer stack is composed by Si/SiO(500)/[Ta(3)/Pt(5)]×3/Ir$_{20}$Mn$_{80}$(8)/Co$_{40}$Fe$_{40}$B$_{20}$(4)/MgO(2,5)/Co$_{40}$Fe$_{40}$B$_{20}$(2)/Ta(0,56)Co$_{85}$Gd$_{15}$(10)Cu(1)Pt(3) where the layer thicknesses are given in nanometer in the brackets. Starting from a Si/SiO(500nm) substrate with SiO made by thermal oxidation, a [Ta(3)/Pt(5)]×3 seed layer is deposited to insure a low access resistance of the bottom electrode and a high (111) crystallographic texture. On top of the seed layer, IrMn(8) is grown and follows the (111) texture. The core of the MTJ is the CoFeB(4)/MgO(2,5)/CoFeB(2)/Ta(0,56) multilayer with amorphous CoFeB electrodes. The active AOS layer is exchange biased though the thin Ta layer and is made of a thick Co$_{85}$Gd$_{15}$(10). This layer, with this thickness and composition, has been studied in [Lin23] and could be reversed using a single laser pulse. Finally a Cu(1)Pt(3) capping layer protects the layer against oxidation. After deposition, the multilayer stack is annealed at 300°C for 1h under an in plane applied field of 600 Oe. The hysteresis loop measured by Vibrating Sample Magnetometry (VSM) is reported in figure 1(a). Well defined and separated reversals could be observed with an exchange bias field of 180 Oe induced by the IrMn on a hard CoFeB layer and a slight ferromagnetic coupling of 6 Oe between the hard and soft layer. The soft CoFeB(2)/Ta(0,56)CoGd(10) layer reverses in a single step, indicating a strong coupling through the Ta layer.

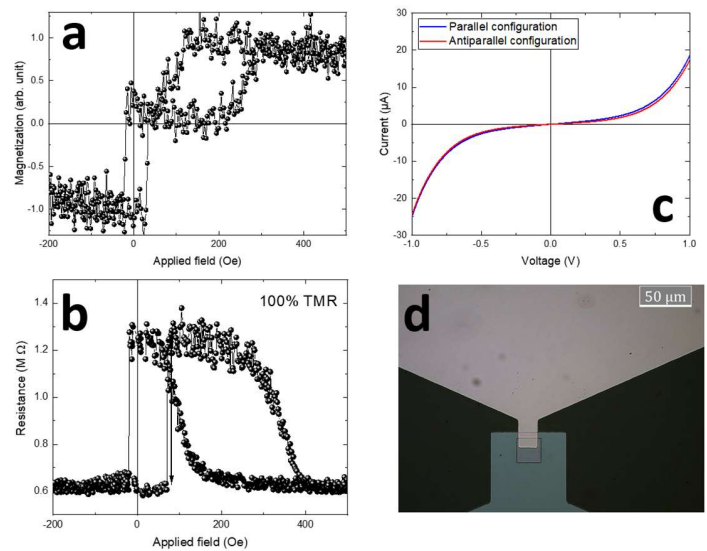

**Figure 1:** Magnetic and transport properties of a Si/SiO(500)/[Ta(3)/Pt(5)]*3/IrMn(8)CoFeB(4)/MgO(2,5)/CoFeB(2)/Ta(0,56)CoGd15%(10)Cu(1)Pt(3) MTJ a) Normalized magnetisation as function of an in-plane applied field, H, using a VSM and (b) corresponding R(H) response .(c) Non linear I(V) curves for both parallel and antiparallel magnetizations configurations. (d) Design of the MTJ electrodes : dark (respectively light) grey are the bottom (respectively top) electrodes.

Finally, a 3 steps lithography process was used to define the MTJ device that allows to applied a voltage across the MgO tunnel barrier (figure 1(d)). The Ti(10)/Au(150) bilayer used as the top electrode for the MTJ does not completely cover the Cu(1)Pt(3) capping layer to allow the laser to shine on the MTJ. It is estimated that approximately 25% of the upper Cu/Pt electrode is covered, which consequently means that the laser does not irradiate that area. The I(V) curves show a tunneling characteristics in both parallel and antiparallel states (figure 1(c)). The $R(H)$ curve shows a TMR value above 100% which is a sign of good crystallization and symmetry filtering of the MgO tunnel barrier (figure 1(b)). Such high value of TMR has not been reach in optically active MTJ with PMA electrodes. It can be noticed that after the lithography process, a stronger ferromagnetic coupling between the electrodes sets up, with a typical value of 31 Oe.

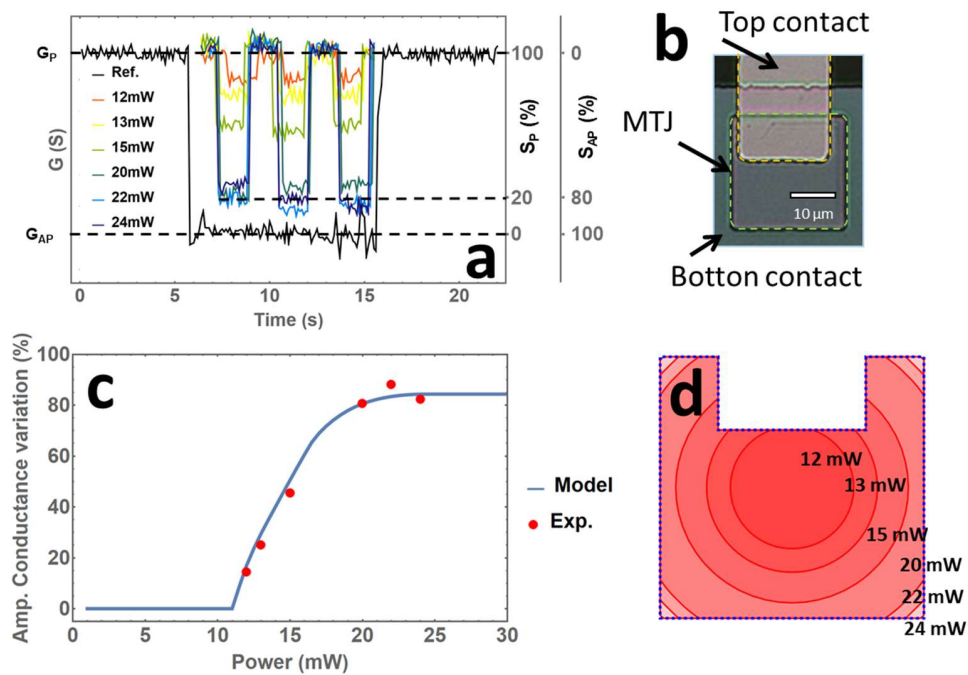

**Figure 2:** All Optical Switching for the IrMn(8)CoFeB(4)/MgO(2,5)/CoFeB(2)/ MTJ (a) Conductance versus time. The black curve is obtained by applying an in-plane magnetic field to reach the parallel and antiparallel configurations. Colour curves are obtained for different laser power ranging from 12 to 24 mW. Starting from parallel configuration, successive laser pulses are applied at time scale around 7, 8.5, 10.5, 12, 13.5 and 15s. (b) Optical microscopy image of the MTJ electrodes configuration. (c) Variation of conductance versus laser power (red dots) and model of surface variation of laser power above power threshold. (d) Corresponding surfaces considering a spot centred on the MTJ.

As a first step, the conductance has been measured as a function of laser power and number of laser pulses. We use a Ti: Sapphire femtosecond-laser source and regenerative amplifier for the pump laser beam for the experiment. Wavelength, pulse duration, and repetition rate of the fs laser were 800 nm, ≈ 35 fs, and 5 kHz, respectively. We utilize a source meter (Keithley 2400) to measure MTJ resistance R by applying a voltage of 30 mV. In figure 2(a), starting from the parallel state (P) obtained by field saturation of both magnetic electrodes, the conductance is measured after laser pulses applied at time scales around 7, 8.5, 10.5, 12, 13.5 and 15s. Considering tunnel conduction in parallel channels with areas where the magnetizations are parallel, $S_P$, (respectively antiparallel, $S_{AP}$) on either side of the tunnel junction, it follows that the conductance of the tunnel barrier is given by $G = G_{AP}\frac{S_{AP}}{S} + G_P\frac{S_P}{S}$ where S is the total surface of the junction. Rewritting this equation leads to $G = G_{AP} + (G_P - G_{AP})\frac{S_P}{S}$ where $G_P - G_{AP}$ is the total variation of conductance when the MTJ undergoes a full reversal from P to AP state. As a result, the level of $G$ gives a direct measure of $\frac{S_P}{S}$ or $\frac{S_{AP}}{S}$. As shown in previous studies [Lin23], the increase in laser power leads to the appearance of a larger inverted domain in CoGd, i.e. a decrease (respectively increase) of $\frac{S_P}{S}$ (respectively $\frac{S_{AP}}{S}$). This situation corresponds to values of $G$ going from $G_P$ towards $G_{AP}$. The decrease of conductance saturates for powers above 20 mW at a value corresponding to $\frac{S_{AP}}{S}$ around 70 to 80%. This corresponds to a full reversal of the part of the tunnel barrier that is not covered by the Ti/Au top electrode. A further increase of the power, above 25 mW leads to the destruction of the MTJ.

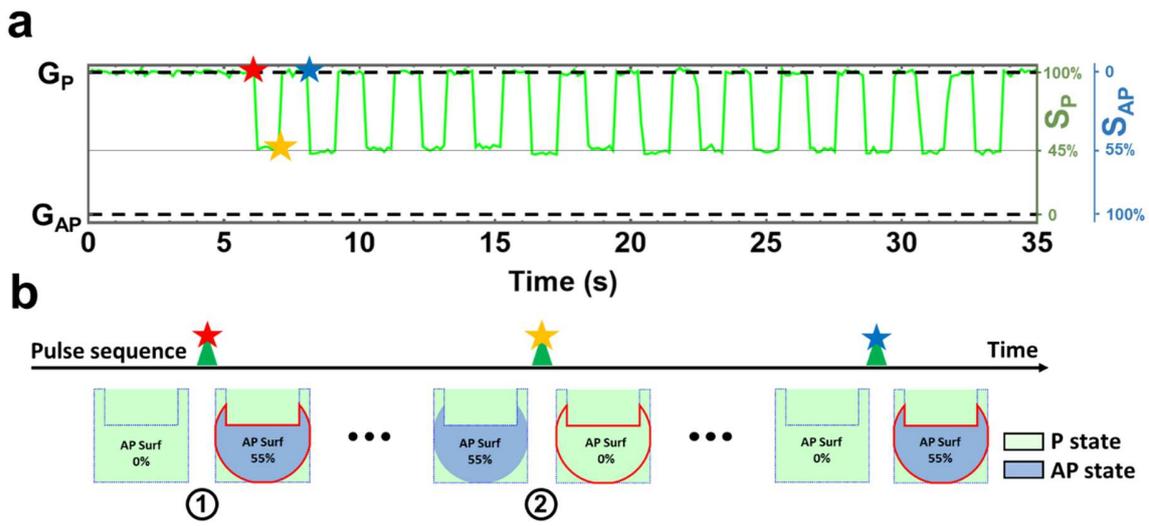

**Figure 3:** (a) Conductance as a function of time. In dashed black, the level of G obtained by applying a magnetic field to induce parallel to antiparallel and antiparallel to parallel configurations. In green, the variation obtained with the laser train of pulses. (b) Sketch of the evolution of the magnetic configuration as a function of time and the pulse sequence of the red, orange and blue pulses. Green (respectively blue) are areas of the MTJ that are in the parallel (respectively antiparallel) configuration of magnetizations. When the laser is applied, the contour of the laser spot hitting the MTJ with sufficient power to reverse the CoFeB/Ta/CoGd layer is plotted in red. The starting configuration is the parallel state. While recording R. the laser pulses are shined on the MTJ every second.

We modelled the spatial distribution of power of the laser beam $P(r)$ using a Gaussian profile $\frac{1}{\sqrt{2\pi}\sigma}e^{-\frac{r^2}{2\sigma^2}}$ where r is the distance to laser spot center which has been considered located at the center of the MTJ. In our model, we fixed the threshold optical power for the reversal of the CoFeB/Ta/CoGd magnetization by AOS to 11 mW. By increasing the power of the laser beam, we calculated the surface of the junction for which power is above 11 mW and should have experienced AOS. We excluded the part of the MTJ that is covered by the top Ti/Au electrode. We then estimated the expected variation of conductance that should be measured. A very good accordance to the experimental measurements could be obtained by using σ = 14 μm (Figure 2(c)).

Still starting from the parallel configuration of magnetizations, we repeated our experiment with a power that results in a variation in conductance corresponding to a magnetic configuration whose surface is 55% anti-parallel and a train of successive laser pulses is applied to the MTJ (figure 3). To obtain 55¨% of the MTJ surface reversed, our simulation using a gaussian profile of power with a power threshold to get reversal shows that a part of the laser beam with sufficient intensity to induce a magnetization reversal hit the Ti/Au electrode. The changes of the magnetic state are reported in figure 3 for 28 successive pulses. After the first laser pulse (figure 3-b1), 55% of the MTJ surface is reversed. . The remaining 45% consists of unirradiated areas (covered by the Ti/Au electrode) and areas where the critical fluence threshold is not reached. This state of conductance is kept up to the next laser pulse for which the MTJ configuration is brought back to fully P state (figure 3-b2). The successive pulses will cause the tunnel junction to transit between these two states. The toggle switching of the CoFeB/Ta/CoGd layer occurs.

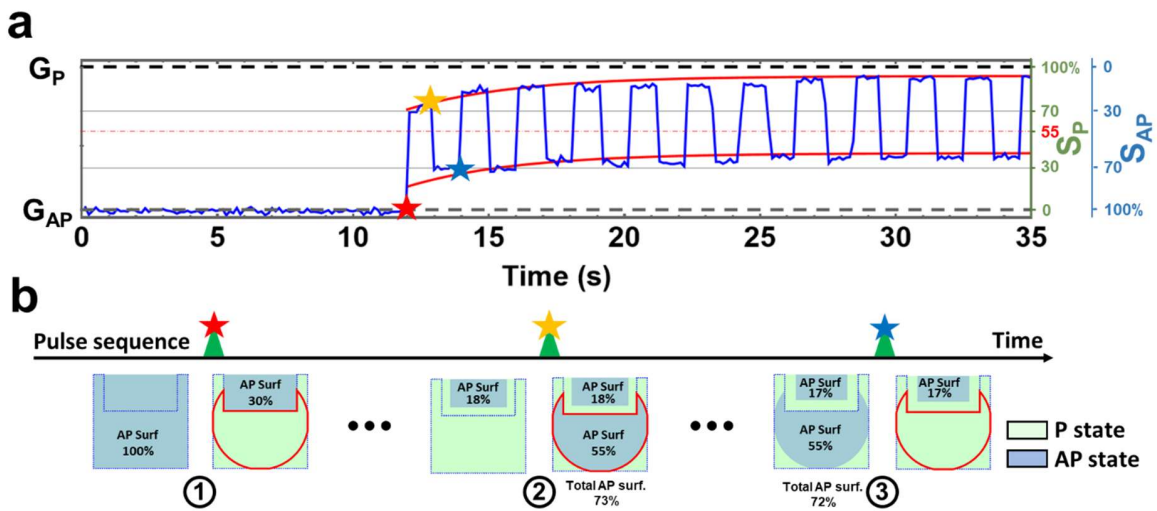

**Figure 4:** (a) Conductance as a function of time. In dashed black, the level of G obtained by applying a magnetic field to induce parallel to antiparallel and antiparallel to parallel configurations. In blue, the variation obtained with the laser train of pulses starting from antiparallel configuration. In red, an exponential decay fit of the conductance variation corresponding to the evolution of the apparent parallel state. (b) Sketch of the evolution of the magnetic configuration as a function of time and pulse sequence. Green (respectively blue) are areas of the MTJ that are in the parallel (respectively antiparallel) configuration of magnetizations. When the laser is applied, the contour of the laser spot hitting the MTJ with sufficient power to reverse the CoFeB/Ta/CoGd layer is plotted in red. The starting configuration is the antiparallel state. While recording R. the laser pulses are shined on the MTJ every second.

The process appears slightly different when the same train of pulses, with the same laser power, is applied to the MTJ but starting from the magnetic AP configuration. The results are reported in figure 4. With respect to the results reported in figure 3, after the first pulse, the MTJ should undergo a transition towards a state where 55% of the MTJ is reversed towards the parallel state (the area delimited by the red line in figure 4-b1 for which the laser has a power above the switching power – dotted red line for the level of conductance). The conductance however shows that the MTJ evolves towards a state that corresponds to 70% of the MTJ is in the parallel state. This corresponds to the reversal of the total surface of the MTJ not covered by the Ti/Au electrode towards the P configuration (Figure 4(b1)). Furthermore, after the laser pulse, the conductance keeps increasing towards the full P conductance value. This can be explained considering that part of the MTJ under the Ti/Au undergoes an AP to P switching (Figure 4(b2)). This evolution could be reproduced using a temporal exponential decay, probably due to some magnetic domain relaxation mechanics. This contrast with the 55% conductance variation observed starting from the P state with a stable magnetic configuration before the next pulse is applied to the MTJ. Both features could be explained by the presence of the ferromagnetic coupling observed in the $R(H)$ loop. Indeed, after the first pulse, the ferromagnetic coupling between the electrodes would not only favor the P alignment in the exposed area but it could also potentially facilitate domain wall growth of P domains within all the heated sections of the MTJ, including below the Ti/Au electrode. Such coupling explains the measured conductance variation of 70% instead of the expected 55%, as well as the observed relaxation. A second laser pulse leads to the 55% conductance decrease of MTJ from the P to AP plus the part of MTJ under the Ti/Au that was in the initial AP state, a total of 73% of the MTJ is in the AP state (Figure 4(b2)), right part of the drawing with the laser on). From the third pulse, the non-covered MTJ part follows the 55% conductance increase, as for the figure 3 and the creep proceeds in the covered region of the MTJ. The behavior starting from full P (figure 3) and full AP (figure 4) states merge in a single one.

## Discussion and conclusion

In this paper, we demonstrate single pulse All Optical Helicity-Independent Switching in in-plane magnetized magnetic tunnel junction. The in plane configuration relaxes all the constrains linked to perpendicular magnetic anisotropy systems and to writing by STT or SOT : much simple multilayer structure, much robust to thermal annealing, much frugal in energy, higher volume of the storage magnetic element and so higher stability over time, lower size of the device and certainly higher writing speed. Furthermore, we extended the TMR ratio above 100% which could not be achieved in MTJ that were written by optical pulses up to now. In our study, the use of an opaque upper electrical electrode due to our technological limitations, does not allow full switching at each pulse and the reasons have been discussed. However, the use of transparent top electrode, like ITO, would solve this technical issue. These new results can potentially help in the development of new magnetic random access memory architectures thanks to the simplicity of their implementation.

## Acknowledgements

We acknowledge financial support from the ANR (ANR-17-CE24-0007 UFO project) and (ANR-23-CE30-0047 Slam Project) and through the France 2030 government grants PEPR Electronic EMCOM (ANR-22- PEEL-0009) and PEPR SPIN (ANR-22-EXSP 0007) and JSPS KAKENHI JP24K22964. The work was supported by the MAT-PULSE-Lorraine Université d'Excellence project (ANR-15-IDEX-04-LUE), the European Cooperation in Science and Technology (COST) Action CA23136 CHIROMAG, the metropole Grand Nancy -Optimag project.

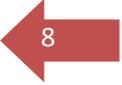